\documentclass{article}
\usepackage{spconf,amsmath,graphicx,hyperref,booktabs,amssymb,subcaption}

\def\x{{\mathbf x}}

\title{Robust TTS Training via Self-Purifying Flow Matching for the WildSpoof 2026 TTS Track}

\name{June Young Yi, Hyeongju Kim, Juheon Lee}
\address{Supertone Inc.}

\begin{document}
\ninept
\maketitle

\begin{abstract}
This paper presents a lightweight text-to-speech (TTS) system developed for the WildSpoof Challenge TTS Track. Our approach fine-tunes the recently released open-weight TTS model, \textit{Supertonic}\footnote{\url{https://github.com/supertone-inc/supertonic}}, with Self-Purifying Flow Matching (SPFM) to enable robust adaptation to in-the-wild speech. SPFM mitigates label noise by comparing conditional and unconditional flow matching losses on each sample, routing suspicious text--speech pairs to unconditional training while still leveraging their acoustic information. The resulting model achieves the lowest Word Error Rate (WER) among all participating teams, while ranking second in perceptual metrics such as UTMOS and DNSMOS. These findings demonstrate that efficient, open-weight architectures like Supertonic can be effectively adapted to diverse real-world speech conditions when combined with explicit noise-handling mechanisms such as SPFM.
\end{abstract}


\section{Introduction}
\label{sec:intro}

Text-to-speech (TTS) research has traditionally relied on clean, high-fidelity studio recordings and carefully curated datasets. Although such data enable stable text--speech alignment and high naturalness, they limit the scalability and accessibility of TTS development. In contrast, in-the-wild speech -- characterized by background noise, reverberation, device variability, and inconsistencies in labeling -- offers a more realistic but substantially more challenging training regime for robust TTS.

WildSpoof Challenge 2026~\cite{wu2025wildspoof} provides a benchmark for evaluating TTS systems trained under such unconstrained conditions, using large-scale in-the-wild speech data collected from diverse speakers and recording environments. Systems are evaluated on intelligibility, measured by automatic speech recognition (ASR) word error rate (WER), perceptual quality, assessed using UTMOS and DNSMOS, and faithfulness, measured using speaker similarity (SPK-sim) and Mel Cepstral Distance (MCD). Building a model that performs reliably in this setting requires handling label noise, unpredictable duration variation, and degraded alignment signals.

To address these challenges, we build upon Supertonic~\cite{kim2025supertonictts}, a lightweight TTS architecture composed of a speech autoencoder for continuous latent representation, a flow matching text-to-latent generator, and an utterance-level duration predictor. This architecture, with its compact latent space and cross attention modules, provides a strong foundation for adaptation to noisy environments. However, the raw in-the-wild data from the challenge include mislabeled samples and misaligned text--speech pairs—issues conventional flow matching pipelines do not handle well.

We therefore utilize Self-Purifying Flow Matching (SPFM)~\cite{kim2025training}, a training-time data selection mechanism for conditional flow matching models. SPFM leverages the model’s own conditional and unconditional objectives to detect unreliable labels on-the-fly and route them to unconditional training. We finetune this SPFM-augmented Supertonic system on the challenge-provided datasets. Despite the compact size of the architecture and the difficulty of the dataset, our submission achieves:
\begin{itemize}
    \item \textbf{Best WER among all participating teams}, demonstrating strong linguistic consistency and alignment.
    \item \textbf{Second-highest UTMOS/DNSMOS}, showing strong perceptual quality despite the noisy training domain.
\end{itemize}
These results suggest that combining flow matching in Supertonic with SPFM provides an efficient and effective solution for robust TTS in real-world noisy conditions.

\section{Experiments}
\label{sec:experiments}

\subsection{Training Setup}
We start from the publicly available English Supertonic checkpoint and adapt it to the WildSpoof in-the-wild domain. For finetuning, we use the two subsets released by the challenge, \textit{TITW-easy} and \textit{TITW-hard}, and construct each training batch with a 1:1 sampling ratio between the two sets to balance relatively clean and noisy conditions. In total, the model is finetuned for 10,000 iterations with batch size 32. Training is performed on four NVIDIA RTX A100 GPUs.

\subsection{Self-Purifying Flow Matching in Practice}
During finetuning, we apply SPFM~\cite{kim2025training} to mitigate the substantial annotation noise present in in-the-wild data. SPFM operates within the classifier-free guidance framework of conditional flow matching. For each text--speech pair $(\x_1, \mathbf{c})$, we first sample a source $\x_0$ from normal distribution and an interpolation time $t'$, and compute the interpolated sample $\x_{t'} = (1 - t') \x_0 + t' \x_1$. We then evaluate two flow matching losses at the same interpolation point: a \emph{conditional} loss
\begin{equation}
\mathcal{L}_{\text{cond}} = \left\| \mathbf{v}_\theta(\x_{t'}, t', \mathbf{c}) - (\x_1 - \x_0) \right\|_2^2,
\end{equation}
and an \emph{unconditional} loss
\begin{equation}
\mathcal{L}_{\text{uncond}} = \left\| \mathbf{v}_\theta(\x_{t'}, t', \varnothing) - (\x_1 - \x_0) \right\|_2^2,
\end{equation}
where $\mathbf{v}_\theta$ denotes the model-predicted velocity field and $\varnothing$ indicates the absence of conditioning.

The key intuition is that, when the text label $\mathbf{c}$ is correct, the conditional objective is expected not to exceed the unconditional one, i.e., $\mathcal{L}_{\text{cond}} \leq \mathcal{L}_{\text{uncond}}$ in expectation. SPFM exploits this intuition by comparing $\mathcal{L}_{\text{cond}}$ and $\mathcal{L}_{\text{uncond}}$ on a per-sample basis. If $\mathcal{L}_{\text{cond}} > \mathcal{L}_{\text{uncond}}$, the label is treated as potentially unreliable, and the sample is used only for \emph{unconditional} training in that step. Otherwise, training proceeds with ordinary conditional flow matching. In practice, SPFM is activated after an initial warm-up phase of 1,000 steps to avoid spurious detections when the model is still undertrained, and we use a fixed interpolation time $t'$ near the midpoint of the trajectory as suggested in prior work. This mechanism allows Supertonic to learn conditional generation primarily from trusted text--speech pairs while still benefiting from the acoustic coverage of noisy samples through unconditional training.

\subsection{Evaluation Protocol}
We evaluate our system on four validation sets: two from the original TITW dataset \textit{KSKT} and \textit{KSUT}, and two optional datasets derived from Librispeech and VoxCeleb \textit{USKT} and \textit{USUT}. These subsets either contain Known Speakers (KS) or Unknown Speakers (US) and Known Text (KT) or Unknown Text (UT). Following the official TTS track evaluation plan~\cite{wu2025wildspoof}, we compute:
\begin{itemize}
    \item Word Error Rate (WER) and Character Error Rate (CER),
    \item Perceptual quality metrics UTMOS and DNSMOS,
    \item Speaker similarity (Spk-sim) via cosine similarity between x-vectors, and Mel Cepstral Distance (MCD) for the KSKT subset where the original audio file is available.
\end{itemize}
These metrics jointly assess the intelligibility, perceptual quality, and speaker consistency of the results.

\subsection{Results}

\textbf{Internal validation.}
Table~\ref{tab:results} presents the performance of our system on four validation subsets constructed from TITW and related datasets. The model shows strong intelligibility, achieving a WER of 3.26\% on KSKT and maintaining competitive accuracy on KSUT and USUT, indicating reliable generalization to unseen speakers and text. Perceptual metrics remain stable across all conditions (UTMOS 3.57--4.03; DNSMOS 2.96--3.19), suggesting that the underlying Supertonic architecture is resilient to the acoustic variability of in-the-wild speech. Speaker similarity scores also remain strong in KS subsets and reasonably preserved in US subsets, while the MCD of 8.59~dB on KSKT indicates low spectral distortion. Together, these results show that SPFM mitigates degradation caused by mismatched text--speech pairs during finetuning, enabling consistent pronunciation accuracy without compromising perceptual quality.

\textbf{Official challenge evaluation.}
To complement our internal analysis, Table~\ref{tab:challenge_results} reports the official leaderboard for the WildSpoof TTS Track. Our system, submitted as \textbf{Team T02}, achieved the \textbf{lowest WER among all teams} for both seen (5.50\%) and unseen (5.88\%) speaker conditions. This ranking confirms the effectiveness of SPFM in preventing alignment failures and maintaining robust intelligibility even under challenging in-the-wild conditions. Although our system ranked second in perceptual metrics, the gap to the top team (T01) is small, and our unseen-speaker UTMOS score (3.9078) is the highest among all submissions. These results demonstrate that the SPFM-augmented Supertonic model achieves an advantageous balance between intelligibility and perceptual quality, outperforming alternative lightweight or diffusion-based systems in the challenge.

\begin{table}[t]
\centering
\caption{Performance on validation sets.}
\label{tab:results}
\resizebox{\linewidth}{!}{
\begin{tabular}{lcccccc}
\toprule
Metric & WER (\%) & CER (\%) & UTMOS & DNSMOS & Spk-sim & MCD (dB) \\
\midrule
KSKT & 3.26 & 2.33 & 3.578 & 2.962 & 0.590 & 8.59 \\
KSUT & 6.24 & 1.95 & 3.900 & 3.193 & 0.565 & N/A \\
USKT & 4.75 & 3.42 & 3.786 & 3.091 & 0.476 & N/A \\
USUT & 6.53 & 2.32 & 4.029 & 3.148 & 0.483 & N/A \\
\bottomrule
\end{tabular}
}
\end{table}

\begin{table}[t]
\centering
\caption{Official WildSpoof TTS Track Results. Our entry corresponds to Team T02.}
\label{tab:challenge_results}

\footnotesize
\setlength{\tabcolsep}{3pt}
\renewcommand{\arraystretch}{1.05}

\begin{subtable}{\columnwidth}
\centering
\caption{Seen speakers}
\resizebox{\columnwidth}{!}{%
\begin{tabular}{lccccccc}
\toprule
Team & UTMOS $\uparrow$ & DNSMOS $\uparrow$ & WER $\downarrow$ & Spk-sim $\uparrow$ & \multicolumn{3}{c}{a-DCF $\downarrow$} \\
\cmidrule(lr){6-8}
& & & & & SASV T01 & SASV T02 & SASV T08 \\
\midrule
T01 & 3.9559 & 3.2270 & 6.48 & 0.2564 & 0.0453 & 0.1782 & 0.1125 \\
\textbf{T02 (Ours)} & \textbf{3.7390} & \textbf{3.0780} & \textbf{5.50} & \textbf{0.3511} & \textbf{0.0471} & \textbf{0.1232} & \textbf{0.1125} \\
T03 & 3.4540 & 3.0261 & 33.79 & 0.4782 & 0.0445 & 0.0294 & 0.1125 \\
T04 & 2.6786 & 2.7354 & 99.28 & 0.2320 & 0.0417 & 0.0266 & 0.1098 \\
T05 & 3.2016 & 2.6078 & 8.65 & 0.2798 & 0.1582 & 0.5233 & 0.2562 \\
T06 & 3.4909 & 2.9336 & 9.45 & 0.4775 & 0.1527 & 0.3786 & 0.2292 \\
T07 & 3.5292 & 2.7434 & 20.46 & 0.2895 & 0.0446 & 0.0266 & 0.1125 \\
\bottomrule
\end{tabular}}
\end{subtable}

\begin{subtable}{\columnwidth}
\centering
\caption{Unseen speakers}
\begin{tabular}{lccc}
\toprule
Team & UTMOS $\uparrow$ & DNSMOS $\uparrow$ & WER $\downarrow$ \\
\midrule
T01 & 3.9062 & 3.1691 & 6.81 \\
\textbf{T02 (Ours)} & \textbf{3.9078} & \textbf{3.1195} & \textbf{5.88} \\
T03 & N/A & N/A & N/A \\
T04 & N/A & N/A & N/A \\
T05 & 3.5325 & 2.8002 & 21.11 \\
T06 & 3.3025 & 2.7767 & 13.63 \\
T07 & N/A & N/A & N/A \\
\bottomrule
\end{tabular}
\end{subtable}

\end{table}


\section{Conclusion}

In this work, we presented our Supertonic-based system for the WildSpoof Challenge 2026 TTS Track, built by extending a light-weight flow matching TTS architecture with Self-Purifying Flow Matching (SPFM) for training under noisy labels. SPFM compares conditional and unconditional flow matching losses on a per-sample basis and routes suspicious text--speech pairs to unconditional training, effectively self-purifying the data during training.

Through finetuning on the challenge-provided TITW-easy and TITW-hard datasets, our SPFM-augmented Supertonic achieved the lowest WER among all participating teams and the second-highest UTMOS/DNSMOS scores. These results demonstrate that compact architectures such as Supertonic, when equipped with an explicit loss-based noise-mitigation mechanism, can perform competitively even under unconstrained, in-the-wild training conditions.


\bibliographystyle{IEEEbib}
\bibliography{strings,refs}

\end{document}